\documentstyle[11pt]{article}
\begin{document}
\begin{center}
{\Large \bf ``Big" Divisor $D3/D7$ Swiss Cheese Phenomenology}
\vskip 0.1in { Aalok Misra\footnote{e-mail: aalokfph@iitr.ernet.in}\\
Department of Physics, Indian Institute of Technology,
Roorkee - 247 667, Uttarakhand, India}
\author{Aalok Misra}
\end{center}
\thispagestyle{empty}
\begin{abstract}
We review progress made over the past couple of years in the field of Swiss Cheese Phenomenology involving a mobile space-time filling $D3$-brane and stack(s) of fluxed $D7$-branes wrapping the ``big" (as opposed to the ``small") divisor in (the orientifold of a) Swiss-Cheese Calabi-Yau. The topics reviewed include reconciliation of large volume cosmology and phenomenology, evaluation of soft supersymmetry breaking parameters, one-loop RG-flow equations' solutions for scalar masses, obtaining fermionic (possibly first two generations' quarks/leptons) mass scales in the ${\cal O}(MeV-GeV)$-regime as well as (first two generations') neutrino masses (and their one-loop RG flow) of around an $eV$. The heavy sparticles and the light fermions indicate the possibility of ``split SUSY" large volume scenario.
\end{abstract}
\section{Introduction and The Setup}
In the context of string compactifications, obtaining $dS$ vacua and realizing the Standard Model have been two important areas of research. In the context of realizing $dS$ vacua,  the complex structure moduli and the axion-dilaton modulus were stabilized with the inclusion of fluxes \cite{fluxesGiddindsetal,Granafluxreview} and the K\"{a}hler moduli could be stabilized only with inclusion of non-perturbative effects. A supersymmetric $AdS$ minimum was obtained in Type IIB orientifold compactification which was uplifted to a non-supersymmetric metastable $dS$ by adding $\overline D3$-brane, in \cite{KKLT}. Subsequently, other uplifting mechanisms were proposed \cite{otherupliftings}. In a different approach with more than one K\"{a}hler modulus  in the context of the Type IIB orientifold compactification in the large volume scenarios (LVS), a non-supersymmetric $AdS$ was realized with the inclusion of perturbative ${\alpha^{\prime}}^3$ correction to the K\"{a}hler potential which was then uplifted to $dS$ vacuum \cite{Balaetal2} a la KKLT. Followed by this, again in the context of Type IIB orientifold compactification in the LVS limit, it was shown in \cite{dSetal} that with the inclusion of (non-)perturbative $\alpha^{\prime}$ corrections to the K\"{a}hler potential and instanton corrections to the superpotential, one can realize {\it non}-supersymmetric metastable $dS$ solution in a more natural way without having to add an uplifting term (via inclusion of $\overline D3$-brane).

From the point of view of embedding (MS)SM  and realizing its matter content from string phenomenology, the questions of supersymmetry breaking and its transmission to the visible sector are among the most challenging issues - the first being controlled mainly by the moduli potentials while the second one by the coupling of supersymmetry-breaking fields to the visible sector matter fields. The breaking of supersymmetry which  is encoded in the soft terms, is supposed to occur in a hidden sector and then  communicated to the visible sector (MS)SM via different mediation processes (e.g. gravity mediation, anomaly mediation, gauge mediation) among which although none is clearly preferred, gravity mediation is the most studied one due to its computational efficiency.

 From the orientifold-limit-of-F-theory point of view \cite{Sen} corresponding to type IIB compactified on a Calabi-Yau three fold $Z$-orientifold with $O3/O7$ planes, one requires a Calabi-Yau four-fold $X_4$ elliptically fibered (with projection $\pi$) over a 3-fold $B_3(\equiv CY_3-$orientifold)  where $B_3$ could either be a Fano three-fold or an $n$-twisted ${\bf CP}^1$-fibration over ${\bf CP}^2$ such that pull-back of the divisors in $CY_3$ automatically satisfy Witten's \cite{Witten} unit-arithmetic genus condition  \cite{DDF}. In the case of the latter, the $B_3$ is given by the following toric data:
$$
\begin{array}{c|ccccc}
&D_1&D_2&D_3&D_4&D_5 \\ \hline
{\bf C}^*&1&1&1&-n&0\\
{\bf C}^*&0&0&0&1&1
\end{array}$$
where the divisors $D_{1,2,3}$ are pullbacks of three lines in ${\bf CP}^2$ and the divisors $D_{4,5}$ are two sections of the fibration. From the point of view of M-theory compactified on $X_4$, the non-perturbative superpotential receives non-zero contributions from $M5$-brane instantons involving wrapping around uplifts {\bf V} to $X_4$ of ``vertical" divisors ($\pi({\rm\bf V})$ is a proper subset of $B_3$) in $B_3$. These vertical divisors are either components of singular fibers or are pull-backs of smooth divisors in $B_3$. There exists a Weierstrass model $\pi_0:{\cal W}\rightarrow B_3$ and its resolution
$\mu: X_4\rightarrow {\cal W}$. For the vertical divisors being components of singular fibers, $B_3$ can be taken to be a ${\bf CP}^1$-bundle over $B_2$ with ADE singularity of the Weierstrass model along $B_2$. From the type IIB point of view, this corresponds to pure Yang-Mills with ADE gauge groups on $D7$-branes wrapping $B_2$; the vertical divisors are hence referred to as ``gauge-type" divisors. The pullbacks of smooth divisors in $B_3$ need not have a gauge theory interpretation - they are hence referred to as ``instanton-type" divisors.

 Writing $J=\xi^1D_1+\xi^2D_5$, where $D_1$ and $D_5=D_4+nD_1$ are divisors dual to the holomorphic curves $C^1=D_1\cdot D_4$ and $C^2=D_1\cdot D_2$ in the Mori Cone (for which
$\int_{C_i}J>0$) such that $\int_{C_{1,2}}J=\xi^{1,2}$, the volume of $B_3$ is given by: vol($B_3$)=$1/6(\xi^1D_1+\xi^2D_5)^3$. Using $D_1^3=0, D_1^2D_5=1, D_1D_5^2=n,D_5^3=n^2$, the volumes of $D_{4,5}$ are given by: vol($D_4=D_5-nD_1$)=$(\frac{\partial}{\partial\xi^2}-n\frac{\partial}{\partial\xi^1})$vol($B_3$)
and vol($D_5$)=$\frac{\partial}{\partial\xi^2}$vol($B_3$). One hence obtains:
vol($D_5$)=$(\xi^1+n\xi^2)^2/2$, vol($D_4$)=$(\xi^1)^2/2,$ and
$${\rm vol}(B_3)=\frac{\sqrt{2}}{3n}\left({\rm vol}_{D_5}^{3/2}-{\rm vol}_{D_4}^{3/2}\right),$$
implysing that $B_3$ is of the ``Swiss Cheese" type wherein the ``big" divisor $D_5$ contributes positively and the ``small" divisor $D_4$ contributes negatively. Also, vol($D_4\cap D_5$)=$(\frac{\partial}{\partial\xi^2}-n\frac{\partial}{\partial\xi^1})\frac{\partial}{\partial\xi^2}$
vol($B_3$)=0  indicating that $D_4(\equiv\Sigma_S)$ and $D_5(\equiv\Sigma_B)$ do not intersect implying that there is no contribution to the one-loop contribution to the K\"{a}hler potential from winding modes corresponding to strings winding non-contractible 1-cycles in the intersection locus corresponding to stacks of intersecting $D7$-branes wrapped around $D_{4,5}$(See \cite{berghack} and section {\bf 2}).

For $n=6$ \cite{DDF}, the $CY_4$ will be the resolution of a Weierstrass model with $D_4$ singularity along the first section and an $E_6$ singularity along the second section. The $CY_3$ turns out to be a unique Swiss-Cheese Calabi Yau - an elliptic fibration over ${\bf CP}^2$ - in ${\bf WCP}^4[1,1,1,6,9]$ given by:
$$x_1^{18} + x_2^{18} + x_3^{18} + x_4^3 + x_5^2 - 18\psi \prod_{i=1}^5x_i - 3\phi x_1^6x_2^6x_3^6 = 0.$$
There is one Big divisor $\Sigma_B:x_5=0$ and one Small divisor $\Sigma_S: x_4=0$; ${\cal V}=\frac{1}{9\sqrt{2}}\left(\tau_B^{\frac{3}{2}}-\tau_S^{\frac{3}{2}}\right).$

The ${\cal N}=1$ moduli space is locally factorizable into a Special K\"{a}hler manifold and a K\"{a}hler manifold derivable from the parent ${\cal N}=2$ special K\"{a}ler and quaternionic manifolds respectively:
$${\cal M}_{{\cal N}=1}={\cal M}^{{\cal N}=1}_{sk}(\subset{\cal M}^{{\cal N}=2}_{sk})\times{\cal M}^{{\cal N}=1}_k(\subset{\cal M}^{{\cal N}=2}_q)$$ \cite{Louis_et_al}, where:
$$K_{sk}=-ln[i\int_{CY_3}\Omega(z^{\tilde{a}})\wedge{\bar \Omega}({\bar z}^{\tilde{a}})],$$ $\tilde{a}=1,...,h^{2,1}_-(CY_3)$.
Defining $$\rho\equiv 1 + t^A\omega_A(\in H^2(CY_3)) - F_A\tilde{\omega}^A(\in H^4(CY_3)) + (2F - t^AF_A){\rm vol}(CY_3),$$ where
$A\equiv(\alpha=1,...,h^{1,1}_+,a=1,...,h^{1,1}_-)$,
$$K_q = -2 ln[\int_{CY_3}e^{-2\phi}\langle\rho,\rho\rangle_{\rm Mukai}]=-2ln[ie^{-2\phi}(2(F-{\bar F})-(F_\alpha+{\bar F}_{\alpha})(t^\alpha-{\bar t}^{\alpha}))]$$
In the large orientifold volume limit:
$$F=-1/3!\kappa_{ABC}t^At^Bt^C - i/2\zeta(3)\chi(CY_3) + i\sum_{\beta\in H_2^-(CY_3,{\bf Z})}n^0_\beta Li_3(e^{ik_at^a}),$$ $k_a=\int_{\beta}\omega_a$ and only $\kappa_{\alpha\beta\gamma}, \kappa_{\alpha bc}$ are non-zero. In the Einstein frame:
$$K = K_{sk} - ln[-i(\tau-{\bar\tau})] - 2 ln [{\cal V}+((\tau-{\bar\tau})/(2i))^{3/2}(2\zeta(3)\chi(CY_3) - 4 Im F_{ws})].$$
Detailed expressions for the (open and closed string sector) K\"{a}hler potential are given in section {\bf 2} below.

In  \cite{dSetal,axionicswisscheese,largefNL_r}, we  addressed some cosmological issues like $dS$ realization, embedding inflationary scenarios and realizing non-trivial non-Gaussianities in the context of type IIB Swiss-Cheese Calabi Yau orientifold in LVS. This has been done with the inclusion of (non-)perturbative $\alpha^{\prime}$-corrections to the K\"{a}hler potential and non-perturbative instanton contribution to the superpotential. Now in this setup, for studying phenomenological issues, we also include  a single mobile space-time filling $D3$-brane  and stacks of $D7$-branes wrapping the ``big" divisor $\Sigma_B$ as well as supporting $D7$-brane fluxes. In order to have $ g_{YM}\sim O(1)$ for supporting (MS)SM, in the previously studied LVS models with fluxed $D7$-branes, it was argued that $D7$-brane had to wrap the ``small" divisor $\Sigma_S$. Unlike this, in our setup we have shown the same to be possible with ``big" divisor $\Sigma_B$ due to the possible competing contributions coming from the Wilson line moduli.
Further, we  considered the rigid limit of wrapping of the $D7$-brane around $\Sigma_B$ (to ensure that there is no obstruction to a chiral matter spectrum) which is effected by considering zero sections of $N\Sigma_B$ and hence there is no superpotential generated due to the fluxes on the world volume of the $D7$-brane (See \cite{Jockers_thesis} and references therein).

Working in the $x_2=1$-coordinate patch, for definiteness, and defining $z_1={x_1}/{x_2},\ z_2={x_3}/{x_2},\ z_3=
{x_4}/{x_2^6}$ and $z_4={x_5}/{x_2^9}$ ($z_4$ for later use) three of which get identified with the mobile $D3$-brane position moduli, the non-perturbative superpotential due to gaugino condensation on a stack of $N$ $D7$-branes wrapping $\Sigma_B$ will be proportional to $\left(1+z_1^{18}+z_2^{18}+z_3^3-3\phi_0z_1^6z_2^6\right)^{\frac{1}{N}}$ (See \cite{Ganor1_2,Maldaetal_Wnp_pref}) , which according to \cite{Ganor1_2}, vanishes as the mobile $D3$-brane touches the wrapped $D7$-brane. To effect this simplification, we will henceforth be restricting the mobile $D3$-brane to the ``big" divisor $\Sigma_B$. It is for this reason that we are justified in considering a single  wrapped $D7$-brane, which anyway can not effect gaugino condensation. As explained in \cite{DDF}, a complete set of divisors lying within the K\"{a}hler cone, need be considered so that the complex-structure moduli-dependent superpotential $W_{cs}\sim W_{ED3}$ - the ED3-instanton superpotential - therefore only ${\cal O}(1)$ $D3$-instanton numbers, denoted by $n^s$ corresponding to wrapping of the $ED3$-brane around the small divisor $\Sigma_S$, contribute. We would hence consider either $W_{cs}\sim W_{ED3}(n^s=1)$ for $W\sim W_{ED3}(n^s=1)$ or $W_{cs}=-W_{ED3}(n^s=1)$ with $W\sim W_{ED3}(n^s=2)$.  Hence, the
superpotential in the presence of an $ED3-$instanton is of the type (See \cite{Ganor1_2,Maldaetal_Wnp_pref,Grimm})
\begin{equation}
\label{eq:W_np}
\hskip-0.3in W \sim \left(1 + z_1^{18} + z_2^{18} + z_3^2 - 3\phi_0z_1^6z_2^6\right)^{n_s}\Theta_{n^s}(\tau,{\cal G}^a)e^{in^sT_s},
\end{equation}
where $n^s=1\ {\rm or}\ 2$, is the $D3$-instanton number, the holomorphic pre-factor
$\left(1 + z_1^{18} + z_2^{18} + z_3^2 - 3\phi_0z_1^6z_2^6\right)^{n_s}$ represents a one-loop
determinant of fluctuations around the $ED3$-instanton due to the modification of the warped volume of $\Sigma_S$ because of the presence of the mobile $D3$-brane and $\Theta_{n^s}(\tau,{\cal G}^a)$
is the holomorphic Jacobi theta function of index $n^s$
defined via $\Theta_{n^s}(\tau,{\cal G}^a)=\sum_{m_a}e^{\frac{i\tau m^2}{2}}e^{in^s {\cal G}^am_a}$
(which encodes the contribution of $D1$-instantons in
an $SL(2,{\bf Z})$-covariant form)where ${\cal G}^a=c^a-\tau{\cal B}^a$ with
${\cal B}^a\equiv b^a - lf^a$,  $f^a$ being the components of  two-form fluxes valued in
$i^*\left(H^2_-(CY_3)\right)$; the immersion map $i$ is defined as:
$i:\Sigma^B\hookrightarrow CY_3$.

Now, in the context of intersecting brane world scenarios \cite{int_brane_SM}, bifundamental leptons and quarks are obtained from open strings stretched between four stacks of $D7$-branes and the adjoint gauge fields correspond to open strings starting and ending on the same $D7$-brane. In Large Volume Scenarios, however,  one considers four stacks of $D7$-branes (the QCD stack of 3 corresponding to $U(3)$, the EW stack of 2 corresponding to U(2) and two single corresponding to $U(1)_Y$ and the EW singlets, just like intersecting brane world models) wrapping $\Sigma_B$ but with different  choices of magnetic $U(1)$ fluxes turned on, on the two-cycles which are non-trivial in the Homology of $\Sigma_B$ and not the ambient Swiss Cheese Calabi-Yau. In the dilute flux approximation, $\alpha_i(M_s)/\alpha_i(M_{EW}), i=SU(3),SU(2),U(1)_Y$, are hence unified. By turning on different $U(1)$ fluxes on, e.g., the $3_{QCD}+2_{EW}$ $D7$-brane stacks in the LVS setup, $U(3_{QCD}+2_{EW})$ is broken down to $U(3_{QCD})\times U(2_{EW})$ and the four-dimensional Wilson line moduli $a_{I(=1,...,h^{0,1}_-(\Sigma_B))}$ and their fermionic superpartners $\chi^I$ that are valued, e.g., in the $adj(U(3_{QCD}+2_{EW}))$ to begin with, decompose into the bifundamentals $(3_{QCD},{\bar 2}_{EW})$ and its complex conjugate, corresponding to the bifundamental left-handed quarks of the Standard Model (See \cite{bifund_ferm}).

\section{Large Volume Cosmo-Pheno Reconciliation and Soft SUSY Breaking Parameters}

In this section, we review our results pertaining to obtaining a $10^{12}GeV$ gravitino in the inflationary era and about an ${\cal O}(1-100)TeV$ gravitino in the present era in a single large volume compactification scheme on one hand, and the possibility of obtaining an ${\cal O}(1)\ g_{YM}$ for gauge theories supported on the ``big" divisor in the Swiss-Cheese Calabi-Yau wrapped by $D7$-branes as well as
evaluation of soft supersymmetry breaking parameters in the context of large volume $D3/D7$ Swiss-Cheese phenomenology on the other hand. For the former, the setup includes only a mobile space-time filling $D3$-brane moving inside the Swiss-Cheese Calabi-Yau and for the latter, stack(s) of fluxed $D7$-brane(s) wrapping the ``big" divisor in the same Swiss-Cheese Calabi-Yau are also included in the setup.

There has been a tension between  LVS cosmology and LVS phenomenology studied so far. The scale required by cosmological/astrophysical experiments is nearly the same order as the GUT scale ($\sim 10^{16}$ GeV) while in LVS phenomenology, the supersymmetry-breaking at $TeV$ scale requires the string scale to be some intermediate scale of the order of $10^{11}$ GeV. In this way there is a hierarchy in scales involved on both sides making it impossible to fulfill both requirements in the same string theory setup. Although LVS limits of Type IIB Swiss-Cheese orientifold compactifications have been exciting steps in the search for  realistic models on both cosmology as well as phenomenology sides, this hierarchy is reflected in LVS setups, as a hierarchy of compactification volume requirement from $ \cal{V}$ $\sim 10^6$ (for cosmology requirement, e.g. see \cite{kaehlerinflation}) to $\cal{V}$ $\sim 10^{14}$ (for phenomenology requirement \footnote{In a recent paper \cite{quevedoftheorysusy}, the authors have realized soft terms $\sim$ TeV with $ {\cal V}\sim O(10^6-10^7)$ in the context of String/F-theory models with SM supported on a del Pezzo surface, but with very heavy gravitino.}, e.g. see \cite{conloncal}) and  the tension has remained unresolved in a single string theoretic setup with the Calabi-Yau volume stabilized at a particular value\footnote {There has been a proposal \cite{tension1}, which involves a small CY volume for incorporating high-scale inflation and then evolves the volume modulus over a long range and finally stabilizes it in the large volume minimum with TeV gravitino mass after inflation.}. The main idea of our proposal \cite{D3_D7_Misra_Shukla} pertaining to the geometric resolution of the tension between LVS cosmology and phenomenology, is the motion of a space-time filling mobile $D3$-brane which dictates the gravitino mass via a holomorphic pre-factor in the superpotential - section of (the appropriate) divisor bundle. The gravitino mass is given by: $m_{\frac{3}{2}}=e^{\frac{K}{2}}W M_p\sim({W}/{{\cal V}})M_p$ in the LVS limit.
Denoting the vevs of the mobile $D3$-brane position moduli by $z_{i,(0)}$, consider fluctuations about the same given by $\delta z_{i,(0)}$. Defining $P(\{z_{i,(0)}\})\equiv 1 + z_{1,(0)}^{18} + z_{2,(0)}^{18} + z_{3,(0)}^2 - 3\phi_0 z_{1,(0)}^6z_{2,(0)}^6$, one obtains:
$W \sim {\cal V}^{\alpha n^s - n^s}\left(1 + \frac{\sum_i a_i\delta z_{i,(0)}}{P(\{z_{i,(0)}\})}\right)^{n^s},
$ where $a_i$'s are some order one factors and one assumes $P(\{z_{i,(0)}\})\sim {\cal V}^\alpha$. This yields $m_{\frac{3}{2}}\equiv e^{\frac{\hat{K}}{2}}|\hat{W}|\sim{\cal V}^{n^s(\alpha - 1) - 1}$
in the LVS limit. In order to be able to obtain a $10^{12} GeV$ gravitino at ${\cal V}\sim 10^6$ in the inflationary epoch, one hence requires: $\alpha = 1$ ($n^s\geq2$ to ensure a metastable dS minimum in the LVS limit - see \cite{dSetal}) and geometrically, $(z_1,z_2,z_3)\sim({\cal V}^{\frac{1}{18}},{\cal V}^{\frac{1}{18}},z_3)$ along the non-singular elliptic curve:
$\psi_0{\cal V}^{\frac{1}{9}}z_3z_4 - z_3^2 - z_4^3 \sim {\cal V}$.
A similar analysis for obtaining a $TeV$ gravitino in the present epoch would require: $\alpha=1-{3}/{2n^s}$, which for $n^s=2$  yields $\alpha={1}/{4}$, and geometrically, $(z_1,z_2,z_3)\sim({\cal V}^{\frac{1}{72}},{\cal V}^{\frac{1}{72}},z_3)$ along the non-singular elliptic curve:
$\psi_0{\cal V}^{\frac{1}{36}}z_3z_4 - (z_3^2 + z_4^3)\sim{\cal V}^{\frac{1}{4}},$ embedded inside the Swiss-Cheese Calabi-Yau.  Taking the small divisor's volume modulus and the Calabi-Yau volume modulus as independent variables, one can show that the volume of the Calabi-Yau can be extremized at one value - $10^6$ - for varying positions of the mobile $D3$-brane (See \cite{D3_D7_Misra_Shukla}). Apart from details given in \cite{D3_D7_Misra_Shukla}, some of the major contributory reasons include (a) the $D3$-brane position moduli enter the Calabi-Yau volume-independent holomorphic prefactor in the superpotential and hence the extremization of the overall potential (proportional to the modulus squared of the prefactor) with respect to the volume modulus, is not influenced by this prefactor, and (b) in consistently taking the large volume limit as done in this paper, the superpotential is independent of the Calabi-Yau volume modulus.

For subsequent discussions pertaining to LVS phenomenology, we will also be including stack(s) of fluxed space-time filling $D7$-brane(s) in our setup.
The main idea then behind realizing $O(1)$ gauge coupling is the competing contribution to the gauge kinetic function (and hence to the gauge coupling) coming from the $D7$-brane Wilson line moduli as compared to the volume of the big divisor $\Sigma_B$, after constructing local (i.e. localized around the location of the mobile $D3$-brane in the Calabi-Yau) appropriate involutively-odd harmonic one-form on the big divisor that lies in $coker\left(H^{(0,1)}_{{\bar\partial},-}(CY_3)\stackrel{i^*}{\rightarrow}
H^{(0,1)}_{{\bar\partial},-}(\Sigma_B)\right)$, the immersion map $i$ being defined as:
$i:\Sigma^B\hookrightarrow CY_3$. This will also entail stabilization of the Wilson line moduli at around $ {\cal V}^{-\frac{1}{4}}$ for vevs of around ${\cal V}^{\frac{1}{36}}$ of the $D3$-brane position moduli, the Higgses in our setup. Extremization of the ${\cal N}=1$ potential, as shown in \cite{D3_D7_Misra_Shukla}, shows that this is indeed true.  This way the gauge couplings corresponding to the gauge theories living on stacks of $D7$ branes wrapping the ``big" divisor $\Sigma_B$ (with different $U(1)$ fluxes on the two-cycles inherited from $\Sigma_B$) will be given by:
$g_{YM}^{-2}=Re(T_B)\sim \mu_3{\cal V}^{\frac{1}{18}}$, $T_B$ being the appropriate ${\cal N}=1$ K\"{a}hler coordinate (\cite{Jockers_thesis} and summarized in (\ref{eq:N=1_coords}) and the relevant text below the same) and $\mu_3$(related to the $D3$-brane tension)$=\pi/\kappa^2\sim\pi/(\alpha^\prime)^2$,
implying a finite (${\cal O}(1)$) $g_{YM}$ for ${\cal V}\sim10^6$.

The fluctuations around the Higgses' vevs ${\cal V}^{\frac{1}{36}}$ and the related extremum value of the Wilson line modulus ${\cal V}^{-\frac{1}{4}}$ are:
\begin{eqnarray}
\label{eq:fluctuations}
& & z_{1,2}={\cal V}^{\frac{1}{36}} + \delta z_{1,2},\nonumber\\
& & {\cal A}_1={\cal V}^{-\frac{1}{4}}+\delta {\cal A}_1.
\end{eqnarray}
The soft supersymmetry parameters are related to the expansion of the K\"{a}hler potential and Superpotential for the open- and closed-string moduli as a power series in the open-string (the ``matter fields") moduli.
In terms of the same, the K\"{a}hler potential for the open and closed string moduli, can be written as a power series in the fluctuations $\delta z_i$ and $\delta {\cal A}_I$ as under:
\begin{eqnarray}
\label{eq:K2}
& & K \left(\left\{\sigma^b,{\bar\sigma}^B;\sigma^S,{\bar\sigma}^S;{\cal G}^a,{\bar{\cal G}}^a;\tau,{\bar\tau}\right\};\left\{z_{1,2},{\bar z}_{1,2};{\cal A}_1,{\bar{\cal A}_1}\right\}\right) = - ln\left(-i(\tau-{\bar\tau})\right) - ln\left(i\int_{CY_3}\Omega\wedge{\bar\Omega}\right)\nonumber\\
 & & - 2 ln\Biggl[a\left(T_B + {\bar T}_B - \gamma K_{\rm geom}\right)^{\frac{3}{2}}-a\left(T_S + {\bar T}_S - \gamma K_{\rm geom}\right)^{\frac{3}{2}} + \frac{\chi}{2}\sum_{m,n\in{\bf Z}^2/(0,0)}
\frac{({\bar\tau}-\tau)^{\frac{3}{2}}}{(2i)^{\frac{3}{2}}|m+n\tau|^3}\nonumber\\
& &  - 4\sum_{\beta\in H_2^-(CY_3,{\bf Z})} n^0_\beta\sum_{m,n\in{\bf Z}^2/(0,0)}
\frac{({\bar\tau}-\tau)^{\frac{3}{2}}}{(2i)^{\frac{3}{2}}|m+n\tau|^3}cos\left(mk.{\cal B} + nk.c\right)\Biggr]\nonumber\\
 & & +\frac{C^{KK\ (1)}_s(z^{\tilde{a}},{\bar z}^{\tilde{a}})\sqrt{\tau_s}}{{\cal V}\left(\sum_{(m,n)\in{\bf Z}^2/(0,0)}\frac{\frac{(\tau-{\bar\tau})}{2i}}{|m+n\tau|^2}\right)} + \frac{C^{KK\ (1)}_b(z^{\tilde{a}},{\bar z}^{\tilde{a}})\sqrt{\tau_b}}{{\cal V}\left(\sum_{(m,n)\in{\bf Z}^2/(0,0)}\frac{\frac{(\tau-{\bar\tau})}{2i}}{|m+n\tau|^2}\right)}\nonumber\\
& & \sim -2 ln\left(\sum_{\beta\in H_2^-(CY_3,{\bf Z})} n^0_\beta(...)\right)  + \left(|\delta z_1|^2 + |\delta z_2|^2 + \delta z_1{\bar\delta z_2} + \delta z_2{\bar\delta z_1}\right)\hat{K}_{z_i{\bar z}_j} + \left((\delta z_1)^2 + (\delta z_2)^2\right)\hat{Z}_{z_iz_j} + c.c\nonumber\\
   & & + |\delta{\cal A}_1|^2\hat{K}_{{\cal A}_1\bar{\cal A}_1} + (\delta {\cal A}_1)^2 \hat{Z}_{{\cal A}_1{\cal A}_1} + c.c + \left(\delta z_1\delta{\bar{\cal A}_1} + \delta z_2\delta{\bar{\cal A}_1} \right)
\hat{K}_{z_i\bar{\cal A}_1} + c.c + (\delta z_1\delta{\cal A}_1 + \delta z_2\delta{\cal A}_1 ) \hat{Z}_{z_i{\cal A}_1} +  c.c. + ....\nonumber\\
& &
\end{eqnarray}
The appropriate ${\cal N}=1$ coordinates: $S,T_\alpha,{\cal G}^a$ in the presence of a single $D3$-brane and a single $D7$-brane wrapping the big divisor $\Sigma^B$ along with $D7$-brane fluxes are given as \cite{Jockers_thesis}:
{\small \begin{eqnarray}
\label{eq:N=1_coords}
& & \hskip-0.4cm S = \tau + \kappa_4^2\mu_7{\cal L}_{A{\bar B}}\zeta^A{\bar\zeta}^{\bar B}, \tau=l+ie^{-\phi},\ {\cal G}^a = c^a - \tau {\cal B}^a,\nonumber\\
& & \hskip-0.4cm T_\alpha=\frac{3i}{2}(\rho_\alpha - \frac{1}{2}\kappa_{\alpha bc}c^b{\cal B}^c) + \frac{3}{4}\kappa_\alpha + \frac{3i}{4(\tau - {\bar\tau})}\kappa_{\alpha bc}{\cal G}^b({\cal G}^c
- {\bar {\cal G}}^c) \nonumber\\
& & \hskip-0.4cm + 3i\kappa_4^2\mu_7l^2C_\alpha^{I{\bar J}}a_I{\bar a_{\bar J}} + \frac{3i}{4}\delta^B_\alpha\tau Q_{\tilde{f}} + \frac{3i}{2}\mu_3l^2(\omega_\alpha)_{i{\bar j}} z^i\bigl({\bar z}^{\bar j}-\frac{i}{2}{\bar z}^{\tilde{a}}({\bar{\cal P}}_{\tilde{a}})^{\bar j}_lz^l\bigr)
\end{eqnarray}}
where the axion-dilaton modulous is shifted by a geometric contribution coming from D7-brane moduli and the complexified divisor volume has three new contributions (the second line in expression of $T_\alpha$) namely coming from Wilson line moduli, internal fluxes turned on two cycles and spacetime filling mobile D3-brane fluctuations. The various terms and symbols used above are:
\begin{itemize}
\item
 $\kappa_4$ is related to four-dimensional Newton's constant, $\mu_3$ and $\mu_7$ are $D3$ and $D7$-brane tensions, $\kappa_{\alpha ab}$'s are triple intersection integers of the CY orientifold, and $c^a$ and $b^a$ are coefficients of RR and NS-NS two forms expanded in odd basis of $H^{(1,1)}_{{\bar\partial},-}(CY)$,
\item
 $ {\cal L}_{A{\bar B}}=\frac{\int_{\Sigma^B}\tilde{s}_A\wedge\tilde{s}_{\bar B}}{\int_{CY_3}\Omega\wedge{\bar\Omega}}$,
$\tilde{s}_A\in H^{(2,0)}_{{\bar\partial},-}(\Sigma^B)$,
\item
fluctuations of $D7$-brane in $CY_3$ normal to $\Sigma^B$ are denoted by $\zeta\in H^0(\Sigma^B,N\Sigma^B)$, i.e., they are the space of global sections of the normal bundle $N\Sigma^B$,
\item
${\cal B}\equiv b^a - lf^a$, where $f^a$ are the components of elements of two-form fluxes valued in $i^*\left(H^2_-(CY_3)\right)$, immersion map is defined as:
$i:\Sigma^B\hookrightarrow CY_3$,
\item
 $C^{I{\bar J}}_\alpha=\int_{\Sigma^B}i^*\omega_\alpha\wedge A^I\wedge A^{\bar J}$, $\omega_\alpha\in H^{(1,1)}_{{\bar\partial},+}(CY_3)$ and $A^I$ forming a basis for $H^{(0,1)}_{{\bar\partial},-}(\Sigma^B)$,
\item
 $a_I$ is defined via a Kaluza-Klein reduction of the $U(1)$ gauge field (one-form) $A(x,y)=A_\mu(x)dx^\mu P_-(y)+a_I(x)A^I(y)+{\bar a}_{\bar J}(x){\bar A}^{\bar J}(y)$, where $P_-(y)=1$ if $y\in\Sigma^B$ and -1 if $y\in\sigma(\Sigma^B)$,
\item
 $z^{\tilde{a}}, \tilde{a}=1,...,h^{2,1}_-(CY_3),$ are $D=4$ complex structure deformations of the CY orientifold 
\item
 $\left({\cal P}_{\tilde{a}}\right)^i_{\bar j}\equiv\frac{1}{||\Omega||^2}{\bar\Omega}^{ikl}\left(\chi_{\tilde{a}}\right)_{kl{\bar j}}$, i.e.,
${\cal P}:TCY_3^{(1,0)}\longrightarrow TCY_3^{(0,1)}$ via the transformation:
$z^i\stackrel{\rm c.s.\ deform}{\longrightarrow}z^i+\frac{i}{2}z^{\tilde{a}}\left({\cal P}_{\tilde{a}}\right)^i_{\bar j}{\bar z}^{\bar j}$,
\item
 $z^i$ are scalar fields corresponding to geometric fluctuations of $D3$-brane inside the Calabi-Yau and defined via: $z(x)=z^i(x)\partial_i + {\bar z}^{\bar i}({\bar x}){\bar\partial}_{\bar i}$, and
\item
 $Q_{\tilde{f}}\equiv l^2\int_{\Sigma^B}\tilde{f}\wedge\tilde{f}$, where $\tilde{f}\in\tilde{H}^2_-(\Sigma^B)\equiv{\rm coker}\left(H^2_-(CY_3)\stackrel{i^*}{\rightarrow}H^2_-(\Sigma^B)\right)$.
\end{itemize}
The closed string moduli-dependent K\"{a}hler potential, includes perturbative (using \cite{BBHL}) and non-perturbative (using \cite{Grimm}) $\alpha^\prime$-corrections as well as the loop corrections (using \cite{berghack,loops}). Written out in (discrete subgroup of) $SL(2,{\bf Z})$(expected to survive orientifolding)-covariant form, the perturbative corrections are proportional to $\chi(CY_3)$ and non-perturbative $\alpha^\prime$ corrections are weighted by $\{n^0_\beta\}$, the genus-zero Gopakumar-Vafa invariants that count the number of genus-zero rational curves $\beta\in H_2^-(CY_3,{\bf Z})$. The loop-contributions (dependent also on the complex structure moduli $z^{\tilde{a}}$) arise from KK modes corresponding to closed string or 1-loop open-string exchange between $D3$- and $D7$-(or $O7$-planes)branes wrapped around the ``s" and ``b" divisors. Note, as shown in the introduction, the two divisors of
${\bf WCP}^4[1,1,1,6,9]$ do not intersect implying that there is no contribution from winding modes corresponding to strings winding non-contractible 1-cycles in the intersection locus corresponding to stacks of intersecting $D7$-branes wrapped around the ``s" and ``b" divisors. One sees from (\ref{eq:K2}) that in the LVS limit, loop corrections are sub-dominant as compared to the perturbative and non-perturbative $\alpha^\prime$ corrections.  In fact, the closed string moduli-dependent contributions are dominated by the genus-zero Gopakumar-Vafa invariants which using Castelnuovo's theory of moduli spaces can be shown to be extremely large for compact projective varieties \cite{Klemm_GV} such as the one used. In (\ref{eq:K2}),
$\hat{K}_{i{\bar j}}\equiv\frac{\partial^2 K \left(\left\{\sigma^b,{\bar\sigma}^B;\sigma^S,{\bar\sigma}^S;{\cal G}^a,{\bar{\cal G}}^a;\tau,{\bar\tau}\right\};\left\{\delta z_{1,2},{\bar\delta}{\bar z}_{1,2};\delta{\cal A}_1,{\bar\delta}{\bar{\cal A}_1}\right\}\right)}{\partial C^i{\bar\partial} {\bar C}^{\bar j}}|_{C^i=0}$
and $\hat{Z}_{ij}\equiv\frac{\partial^2 K \left(\left\{\sigma^b,{\bar\sigma}^B;\sigma^S,{\bar\sigma}^S;{\cal G}^a,{\bar{\cal G}}^a;\tau,{\bar\tau}\right\};\left\{\delta z_{1,2},{\bar\delta}{\bar z}_{1,2};\delta{\cal A}_1,{\bar\delta}{\bar{\cal A}_1}\right\}\right)}{\partial C^i\partial C^j}|_{C^i=0}$ - the matter field fluctuations denoted by $C^i\equiv \delta z_{1,2},\delta\tilde{\cal A}_1$. As was shown in \cite{D3_D7_Misra_Shukla}, the following basis of fluctuations simultaneously diagonalizes $\hat{K}_{i{\bar j}}$ and $Z_{ij}$:
\begin{eqnarray}
\label{eq:diagonal}
& & \delta\tilde{{\cal A}_1}\equiv (\beta_1\delta z_1 + \beta_2\delta z_2){\cal V}^{-\frac{8}{9}} +
\delta{\cal A}_1,\nonumber\\
& & \delta{\cal Z}_i\equiv \delta z_i + \lambda_2\delta{\cal A}_1{\cal V}^{-\frac{8}{9}}.
\end{eqnarray}
Hence, for evaluating the physical $\mu$ terms, Yukawa couplings, etc., we will work with the fluctuation fields as given in (\ref{eq:diagonal}).

The superpotential can similarly  be expanded about $\langle z_i\rangle\sim{\cal V}^{\frac{1}{36}}$
and ${\cal A}_0\sim{\cal V}^{-\frac{1}{4}}$, in the basis (\ref{eq:diagonal}) as:
 and these values as:
\begin{eqnarray}
\label{eq:W_exp}
& & W\sim{\cal V}^{\frac{n^s}{2}}\Theta_{n^s}(\tau,{\cal G}^a)e^{in^sT(\sigma^S,{\bar\sigma^S};{\cal G}^a,{\bar{\cal G}^a};\tau,{\bar\tau})}[1 + (\delta {\cal Z}_1 + \delta {\cal Z}_2)\{n^s{\cal V}^{-\frac{1}{36}} + (in^s\mu_3)^3{\cal V}^{\frac{1}{36}}\}\nonumber\\
& & +\delta\tilde{{\cal A}}_1\{-[\lambda_1+\lambda_2](in^s\mu_3){\cal V}^{-\frac{31}{36}} - n^s[\lambda_1+\lambda_2]{\cal V}^{-\frac{11}{12}}\}]+ ((\delta {\cal Z}_1)^2 + (\delta {\cal Z}_2)^2)\mu_{{\cal Z}_i{\cal Z}_i}  + \delta {\cal Z}_1\delta {\cal Z}_2\mu_{{\cal Z}_1{\cal Z}_2}\nonumber\\
& & \hskip-1in+
(\delta\tilde{{\cal A}}_1)^2\mu_{\tilde{\cal A}_I\tilde{\cal A}_I} + \delta {\cal Z}_1\delta\tilde{{\cal A}}_1\mu_{{\cal Z}_1\tilde{\cal A}_I} + \delta {\cal Z}_2\delta\tilde{{\cal A}}_1\mu_{{\cal Z}_2\tilde{\cal A}_I} + ((\delta {\cal Z}_1)^3 + (\delta {\cal Z}_2)^3)Y_{{\cal Z}_i{\cal Z}_i{\cal Z}_i} + ((\delta {\cal Z}_1)^2\delta {\cal Z}_2 + (\delta {\cal Z}_2)^2\delta {\cal Z}_1)Y_{{\cal Z}_i{\cal Z}_i{\cal Z}_j}\nonumber\\
& & + (\delta {\cal Z}_1)^2\delta{\tilde{\cal A}}_1Y_{{\cal Z}_i{\cal Z}_i\tilde{\cal A}_I} + \delta {\cal Z}_1(\delta\tilde{\cal A}_I)^2Y_{{\cal Z}_i\tilde{\cal A}_I\tilde{\cal A}_I} + \delta {\cal Z}_1\delta {\cal Z}_2\delta\tilde{{\cal A}}_1Y_{{\cal Z}_1{\cal Z}_2\tilde{\cal A}_I} + (\delta\tilde{\cal A}_I)^3Y_{\tilde{\cal A}_I\tilde{\cal A}_I\tilde{\cal A}_I}+
 ....,
\end{eqnarray}
where $\sigma^S$ (and for later uses, $\sigma^B$) is the volume of the small (big) divisor complexified by the RR four-form axion, the constants $\lambda_{1,2}$ are functions of the extremum values of the closed string moduli and the Fayet-Illiopoulos parameters corresponding to the NLSM (the IR limit of the GLSM) for an underlying ${\cal N}=2$ supersymmetric gauge theory whose target space is the (toric) projective variety we are considering in our work, and the mobile $D3$-brane position moduli fluctuations $\delta{\cal Z}_i$'s and the $D7-$brane Wilson line moduli fluctuations $\delta\tilde{\cal A}_I$'s ($I$ indexes $dim\left(H^{(0,1)}_{{\bar\partial},-}(CY_3)\right)$ and for simplicity we take $I=1$ corresponding to the harmonic one-form referred to earlier on in this section and explicitly constructed in \cite{D3_D7_Misra_Shukla}), obtained in \cite{D3_D7_Misra_Shukla}, diagonalize the open string moduli metric.

Further, the complete K\"{a}hler potential will consist of the following contribution
 $- 2 ln[a(T_B + {\bar T}_B - \gamma K_{\rm geom})^{\frac{3}{2}}
 -a(T_S + {\bar T}_S - \gamma K_{\rm geom})^{\frac{3}{2}} + ...]$ where $\gamma=\kappa_4^2 T_3$, $T_3$ being the $D3-$brane tension, and hence the same requires evaluation of the geometric K\"{a}hler potential $K_{\rm geom}$. Using GLSM techniques and the toric data for the given Swiss-Cheese Calabi-Yau, the geometric K\"{a}hler potential for the divisor ${\Sigma_B}$ (and ${\Sigma_S}$) in the LVS limit was evaluated in \cite{D3_D7_Misra_Shukla} in terms of derivatives of genus-two Siegel theta functions as well as
two Fayet-Iliopoulos parameters corresponding to the two $C^*$ actions in the  two-dimensional ${\cal N}=2$ supersymmetric gauge theory whose target space is our toric variety Calabi-Yau, and a parameter $\zeta$ encoding the information about the $D3-$brane position moduli-independent (in the LVS limit) period matrix of the hyperelliptic curve $w^2=P_{\Sigma_B}(z)$, $P_{\Sigma_B}(z)$ being the defining hypersurface for $\Sigma_B$. The geometric K\"{a}hler potential for the divisor $D_5$ in the LVS limit is hence given by:
\begin{eqnarray}
\label{eq:Kaehler_D_5}
& & \hskip -0.8in K|_{D_5} =  r_2 - {\left[r_2 - \left(1+|z_1|^2+|z_2|^2\right)\left(\frac{\zeta}{
r_1|z_3|^2}\right)^{\frac{1}{6}}\right]}{4\sqrt{\zeta}}/{3}+|z_3|^2\left\{{\left[r_2 - \left(1+|z_1|^2+|z_2|^2\right)\left(\frac{\zeta}{
r_1|z_3|^2}\right)^{\frac{1}{6}}\right]}{\sqrt{\zeta}}/{3\sqrt{r_1|z_3|^2}}\right\}^2\nonumber\\
& & - r_1 ln\left\{{\left[r_2 - \left(1+|z_1|^2+|z_2|^2\right)\left({\zeta}{
r_1|z_3|^2}\right)^{\frac{1}{6}}\right]}\sqrt{\zeta}/{3\sqrt{r_1|z_3|^2}}\right\}
-r_2 ln\left[\left({\zeta}/{r_1|z_3|^2}\right)^{\frac{1}{6}}\right] \sim{{\cal V}^{\frac{2}{3}}}/{\sqrt{ln {\cal V}}}.
\end{eqnarray}

With a (partial) cancelation between the volume of the ``big" divisor and the Wilson line
contribution (required for realizing $\sim O(1) g_{YM}$ in our setup), in \cite{D3_D7_Misra_Shukla}, we calculated in the large volume limit: (a) using the following expression for the open-string moduli masses:
\begin{eqnarray}
\label{eq:matter_masses_1}
& & m_i^2 = m_{\frac{3}{2}}^2 + V_0 - F^m{\bar F}^{\bar n}\partial_m{\bar\partial}_{\bar n}ln \hat{K}_{i{\bar i}},
\end{eqnarray}
wherein the closed-string moduli $F^m$'s defined via $F^m=e^{\frac{\hat{K}}{2}}\hat{K}^{m{\bar n}}{\bar D}_{\bar n}{\bar W}$, were hence found to be:
$F^{\sigma^B}\sim{\cal V}^{-\frac{n^s}{2}-\frac{17}{18}}; F^{\sigma^S}\sim n^s{\cal V}^{-\frac{n^s}{2}+\frac{1}{36}};F^{{\cal G}^a}\sim n^s{\cal V}^{-\frac{n^s}{2} - 1}$,
and consequently the $D3-$brane position moduli masses were found to be $\sim {\cal V}^{\frac{19}{36}}m_{{3}/{2}}$ and
Wilson line moduli masses were found to be  $ \sim {\cal V}^{\frac{73}{72}}m_{{3}/{2}}$,
(c) the $\mu$ (defined in (\ref{eq:W_exp})) and the
physical $\hat{\mu}$ parameters defined via:
$\hat{\mu}_{ij}={(\frac{{\bar{\hat{W}}}e^{\frac{\hat{K}}{2}}}{|\hat{W}|}\mu_{ij} + m_{\frac{3}{2}}Z_{ij}\delta_{ij} - {\bar F}^{\bar m}{\bar\partial}_{\bar m}Z_{ij}\delta_{ij})}/{\sqrt{\hat{K}_{i{\bar i}}\hat{K}_{j{\bar j}}}},$
(d) the Yukawa couplings $Y_{ijk}$ (defined in
(\ref{eq:W_exp})), the physical Yukawa couplings $\hat{Y}_{ijk}$ defined via:
$\hat{Y}_{ijk}={e^{\frac{\hat{K}}{2}}Y_{ijk}}/{\sqrt{\hat{K}_{i{\bar i}}\hat{K}_{j{\bar j}}\hat{K}_{k{\bar k}}}},$
the $A_{ijk}$-terms defined via
$A_{ijk}=[\hat{K}_m + \partial_m ln Y_{ijk} - \partial_m ln(\hat{K}_{i{\bar i}}\hat{K}_{j{\bar j}}\hat{K}_{k{\bar k}})]$
 and
 (e) the $\hat{\mu}B$-parameters defined through an involved expression (See \cite{conloncal}) involving $\hat{K}_{i{\bar j}}$ and $\hat{Z}_{ij}$.
The soft SUSY parameters we obtained are summarized in Table 1.

\begin{table}[htbp]
\centering
\begin{tabular}{|l|l|}
\hline
Gravitino mass &  $ m_{\frac{3}{2}}\sim{\cal V}^{-\frac{n^s}{2} - 1}$ \\
Gaugino mass & $ M_{\tilde g}\sim m_{\frac{3}{2}}$\\ \hline
$D3$-brane position moduli  & $ m_{{\cal Z}_i}\sim {\cal V}^{\frac{19}{36}}m_{\frac{3}{2}}$ \\
(Higgs) mass & \\
Wilson line moduli mass & $ m_{\tilde{\cal A}_1}\sim {\cal V}^{\frac{73}{72}}m_{\frac{3}{2}}$\\ \hline
A-terms & $A_{pqr}\sim n^s{\cal V}^{\frac{37}{36}}m_{\frac{3}{2}}$\\
& $\{p,q,r\} \in \{{{\tilde{\cal A}_1}},{{\cal Z}_i}\}$\\
\hline
Physical $\mu$-terms & $\hat{\mu}_{{\cal Z}_i{\cal Z}_j}$ \\
 (Higgsino mass) & $\sim{\cal V}^{\frac{37}{36}}m_{\frac{3}{2}}$\\
\hline
Physical Yukawa couplings&
$\hat{Y}_{\tilde{\cal A}_1^2{\cal Z}_i}\sim {\cal V}^{-\frac{127}{72}}m_{\frac{3}{2}}$\\
\hline
Physical $\hat{\mu}B$-terms & $\left(\hat{\mu}B\right)_{{\cal Z}_1{\cal Z}_2}\sim{\cal V}^{\frac{37}{18}}m_{\frac{3}{2}}^2$\\
\hline
\end{tabular}
\caption{Results on Soft SUSY Parameters Summarized}
\end{table}

\section{(RG Flow of) Scalar and Fermionic Masses}

The RG-evolution of various soft-parameters are encoded in a set of coupled differential equations which involve the $\beta$-functions and the anomalous dimensions ($\gamma$'s) of various soft-parameters. The RG flow equations upto two-loops for the ratio of gaugino mass to the square of gauge coupling ($M_a/{g_a}^2$) are given as \cite{sparticlesreview}:
\begin{eqnarray}
& & \frac{dg_a}{dt} = \frac{{g_a}^3}{16 {\pi}^2} b^a + \frac{{g_a}^3}{({16 {\pi}^2})^2}\left[\sum^3_{b=1}B^{(2)}_{ab}{g_b}^2- \sum_{x=u,d,e,\nu} {C^a_x} Tr[{Y_x}^{\dagger}{Y_x}]\right] \nonumber\\
& & \frac{dM_a}{dt} = \frac{2{g_a}^2}{16 {\pi}^2} b^a M_a + \frac{2{g_a}^2}{({16 {\pi}^2})^2}\left[\sum^3_{b=1}B^{(2)}_{ab}{g_b}^2 (M_a+M_b)+ \sum_{x=u,d,e,\nu} {C^a_x}\{Tr[{Y_x}^{\dagger}{\tilde{A_x}}]-M_a Tr[{Y_x}^{\dagger}{Y_x}]\}\right]\nonumber\\
\end{eqnarray}
where $t=2 ln(\frac{{Q_{Q}}}{Q_{EW}})$ defined in terms of ${Q_{EW}}$ which is the phenomenological low energy scale (of interest) and $Q_0$ some high energy scale along with the MSSM gauge coupling $\beta$-functions' given as $b^a=\{{\frac{33}{5},1,-3}\}$, $B^{(2)}_{ab}$ and $C^a_x$ being $3\times3$ and $4\times3$ matrices with ${\cal O}(1-10)$ components and ${\tilde{A_x}}$, $Y_x$ are trilinear A-term and Yukawa-coupling respectively. Further, the first term in each of above equations represents one-loop effect while other terms in the parentheses are two-loop contributions to their RG running implying that $ {\frac{d}{dt}}\left[{\frac{M_a}{{g_a}^2}}\right]=0$ at one-loop.

RG-flow equations of first family of squark and slepton masses results in the following set of equations  which represents the difference in scalar mass-squared values between a low energy scale ${Q_{EW}}$ and a high energy scale $Q_0$ at one-loop level {\cite{Martinreview,QuevedoLHC}}:
\begin{eqnarray}
\label{eq:RGsparticles}
& & {m^2_{\tilde{d_L}}}\bigg|_{Q_{EW}}-m^2_{\tilde{d_L}}\bigg|_{Q_0}={\cal K}_3+{\cal K}_2+{\frac{1}{36}}{\cal K}_1+ \tilde{\Delta}_{\tilde{d_L}}\nonumber\\
& & {m^2_{\tilde{u_L}}}\bigg|_{Q_{EW}}-m^2_{\tilde{u_L}}\bigg|_{Q_0}={\cal K}_3+{\cal K}_2+{\frac{1}{36}}{\cal K}_1+ \tilde{\Delta}_{\tilde{u_L}}\nonumber\\
& & {m^2_{\tilde{d_R}}}\bigg|_{Q_{EW}}-m^2_{\tilde{d_R}}\bigg|_{Q_0}={\cal K}_3+{\frac{1}{9}}{\cal K}_1+ \tilde{\Delta}_{\tilde{d_R}}\nonumber\\
& & {m^2_{\tilde{u_R}}}\bigg|_{Q_{EW}}-m^2_{\tilde{u_R}}\bigg|_{Q_0}={\cal K}_3+{\frac{4}{9}}{\cal K}_1+ \tilde{\Delta}_{\tilde{u_R}}\nonumber\\
& & {m^2_{\tilde{e_L}}}\bigg|_{Q_{EW}}-m^2_{\tilde{e_L}}\bigg|_{Q_0}={\cal K}_2+{\frac{1}{4}}{\cal K}_1+ \tilde{\Delta}_{\tilde{e_L}}\nonumber\\
& & {m^2_{\tilde{e_R}}}\bigg|_{Q_{EW}}-m^2_{\tilde{e_R}}\bigg|_{Q_0}={\cal K}_1+ \tilde{\Delta}_{\tilde{e_R}}\nonumber\\
& & {m^2_{\tilde{\nu}}}\bigg|_{Q_{EW}}-m^2_{\tilde{\nu}}\bigg|_{Q_0}={\cal K}_2+{\frac{1}{36}}{\cal K}_1+ \tilde{\Delta}_{\tilde{\nu}}
\end{eqnarray}
where the parameters ${\cal K}_a$ are defined through the integral (\ref{eq:Kintegrals}) and the difference in the coefficients of ${\cal K}_1$ in the above set of RG equations is due to various weak hyper charge squared values for each scalar.
\begin{eqnarray}
\label{eq:Kintegrals}
& & {\cal K}_a\sim {{\cal O}\bigg({\frac{1}{10}}\biggr)}\int_{ln Q_0}^{ln Q_{\rm EW}} dt g_a^2(t)M_a^2(t)\equiv {{\cal O}\bigg({\frac{1}{10}}\biggr)} \bigg(\frac{M_a}{g_a^2}\bigg)^2\bigg|_{Q_0}\bigg[g_a^4\bigg|_{Q_{EW}}-g_a^4\bigg|_{Q_0}\bigg]\Bigg|_{\rm 1-loop}
\end{eqnarray}
Further, ${\tilde{\Delta}_{\tilde{x}}}$(appearing in (\ref{eq:RGsparticles}))$\sim M_Z^2$, where ${\tilde{x}} \in \{{\tilde{d_L}},{\tilde{d_R}},{\tilde{u_L}},{\tilde{u_R}},{\tilde{e_L}},{\tilde{e_R}},{\tilde{\nu}}\}$ (i.e. the first family of squarks and sleptons)  are D-term contributions which are ``hyperfine" splitting in squark and slepton masses arising due to quartic interactions among squraks and sleptons with Higgs.  Now, in our setup $Q_0\equiv{M_{\rm string}}={M_{\rm GUT}}/{10}\sim 10^{15}GeV$ and $Q_{\rm EW}\sim TeV$. As argued in \cite{Kap_Louis}, the gauge couplings run as follows (up to one loop):
\begin{eqnarray}
\label{eq:RG_flow_1}
\frac{16\pi^2}{g_a^2(Q_{EW})}=\frac{16\pi^2}{g_a^2(Q_0)} + 2b_a ln\biggl[\frac{Q_0}{m_{{3}/{2}}}\biggr] + 2b_a^\prime ln\biggl[\frac{m_{{3}/{2}}}{Q_{EW}}\biggr] + \Delta_a^{\rm 1-loop},
\end{eqnarray}
where $b_a, b_a^\prime$ are group-theoretic factors and $\Delta_a^{\rm 1-loop}\sim {\bf Tr} {ln ({{\cal M}}/{m_{{3}/{2}}})},$ and the mass matrix ${\cal M}$, in our setup, corresponds to the $D7$ Wilson-line modulus ${\cal A}_1$'s mass. Now, ${\cal M}\equiv e^K\hat{K}^{-\frac{1}{2}}\mu^\dagger(\hat{K}^{-1})^T\mu\hat{K}^{-\frac{1}{2}}$ (See \cite{Kap_Louis}). One can show that ${\cal M}_{{\cal A}_1}\sim{\cal V}^{-\frac{13}{6}}$ and  $m_{{3}/{2}}\sim{\cal V}^{-\frac{n^s}{2}-1}$. Hence, for $m_{{3}/{2}}\sim10TeV$ (which can be realized in our setup - see \cite{D3_D7_Misra_Shukla}), one obtains $K_a\sim0.3(TeV)^2$ to be compared with $0.5(TeV)^2$ as obtained in \cite{QuevedoLHC}; an mSUGRA point on the ``SPS1a slope" has a value of around $(TeV)^2$. Further the ${\tilde{\Delta}_{\tilde{x}}}$ contributions, being proportional to $m^2_Z$ and $T_{3{\tilde{x}}}$, $Q_{{\tilde{x}}}$ being fractions (eg. $T_{3{\tilde{u_L}}}={\frac{1}{2}}$, $T_{3{\tilde{d_L}}}=-{\frac{1}{2}}$, $T_{3{\tilde{u_R}}}=0$ and  $Q_{{\tilde{u_L}}}=\frac{2}{3}$, $Q_{{\tilde{d_L}}}=-{\frac{1}{3}}$, $Q_{{\tilde{u_R}}}=-{\frac{2}{3}}$ etc.), are suppressed as compared to ${\cal K}_a$-integrals at one-loop. Given that $\hat{Y}_{\tilde{\cal A}_1^2{\cal Z}_i}\sim {\cal V}^{-\frac{127}{72}}m_{\frac{3}{2}}$, one conjectures that the first two generations' quarks/leptons could have as a representative, the Wilson line modulus ${\cal A}_1$. As per the set of RG-equations (\ref{eq:RGsparticles}), that for the running of first two families of sparticles content, the mass-squared values at electroweak scale can differ from those at the string scale (which is $M_s\sim 10^{15}$ GeV for our case) by a factor of order $(TeV)^2$ with gravitino being $10 TeV$ for Calabi-Yau volume ${\cal V}\sim10^6$. Hence, even at the EW scale, we get very heavy sparticles (squarks/sleptons).

The fermionic bilinear  mass term in the four dimensional effective action can be schematically written in terms of canonically normalized superfields ${\cal Z}^i$ and ${\cal A}^I$  as:
$\int d^4xd^2\theta \,\hat{Y}_{iIJ} {\cal Z}^i {\cal A}^I
{\cal A}^J$.   Now the fermionic masses are generated through Higgs mechanism by giving VEV to Higgs fields:
$M_{IJ} = {\hat {Y}_{iIJ} <z_i }>$
where $\hat {Y}_{iIJ}$'s are
 ``Physical Yukawas" defined as $\hat {Y}_{iIJ}= \frac{e^{\hat{K}/2} Y_{iIJ}}
 {\sqrt {K_{i{\bar i}}} \sqrt {K_{I{\bar I}}} \sqrt {K_{J{\bar J}}}}$ and the Higgs fields
 $z_i$'s are given a vev: $<z_i> \sim {{\cal V}^{\frac{1}{36}}} M_p$
\cite{D3_D7_Misra_Shukla}. 
Next, we discuss the possibility of realizing fermion masses in the range from fraction of an $MeV$ up to a $GeV$  in our setup, possibly corresponding to any of the first two generations' quark/lepton masses.
Using (\ref{eq:W_exp}) the physical Standard Model-like ${\cal Z}_i{\cal A}_I^2$ Yukawa couplings  are found to be (as also evaluated in \cite{D3_D7_Misra_Shukla})
$ \hat{Y}_{{\cal A}_1{\cal A}_1{\cal Z}_i}\sim{\cal V}^{-\frac{199}{72}-\frac{n^s}{2}}$
 \cite{Sparticles_Misra_Shukla}. The leptonic/quark mass is given by:
${\cal V}^{-\frac{199}{72}-\frac{n^s}{2}}$ in units of $M_p$, which  implies a range of fermion mass $m_{\rm ferm}\sim{\cal O}({\rm MeV-GeV})$ for Calabi Yau volume ${\cal V}\sim {\cal O}(6\times10^5-10^6)$. For example, a mass of $0.5$ MeV could be realized with Calabi Yau volume ${\cal V}\sim 6.2\times 10^5, n^s=2$. In MSSM/2HDM models, up to one loop, the leptonic (quark) masses do not change (appreciably) under an RG flow from the intermediate string scale down to the EW scale (See \cite{Das_Parida}). This way, we show the possibility of realizing all fermion masses of first two generations in our setup. Although we do not have sufficient field content to identify all first two families' fermions, we believe that the same could be realized after inclusion of more Wilson line moduli in the setup. The above results also make the possible identification of Wilson line moduli with squarks and sleptons of first two families  \cite{Sparticles_Misra_Shukla}, more robust.

The non-zero Majorana neutrino masses are generated through the Weinberg-type dimension-five operators arising from a lepton number violating term written out schematically as: $$\int d^4x\int d^2\theta e^{\hat{K}/2}\times\left({\cal Z}^2{\cal A}_I^2\in\frac{\partial^2W}{\partial Z^2}{\cal A}_I^2\right)$$ [where ${\cal A}_I=a_I+\theta \chi^I+...$]yielding:
$m_{\nu}={v^2 sin^2\beta \hat{{\cal O}}_{{\cal Z}_i{\cal Z}_j{\cal Z}_k{\cal Z}_l}}/{2M_p}$
 where $\hat{{\cal O}}_{{\cal Z}_i{\cal Z}_i{\cal Z}_i{\cal Z}_i}$ is the coefficient of the physical/noramalized term quartic in the $D3-$brane position moduli ${\cal Z}_i$ which are defined in terms of diagonal basis of K\"{a}hler potential in ${\cal Z}_i$'s and is given as
$$\hat{{\cal O}}_{{\cal Z}_i{\cal Z}_i{\cal Z}_i{\cal Z}_i}={\frac{{e^\frac{\hat{K}}{2}}{\cal O}_{{\cal Z}_i{\cal Z}_j{\cal Z}_k{\cal Z}_l}}{{\sqrt{\hat{K}_{{\cal Z}_i{\bar{\cal Z}}_{\bar i}}\hat{K}_{{\cal Z}_j{\bar{\cal Z}}_{\bar j}}\hat{K}_{{\cal Z}_k{\bar{\cal Z}}_{\bar k}}\hat{K}_{{\cal Z}_l{\bar{\cal Z}}_{\bar l}}}}}},$$ and
$v sin\beta$ is the vev of the $u$-type Higgs $H_u$ with $sin\beta$ defined via
$tan\beta={\langle H_u\rangle}/{\langle H_d\rangle}$.

Now expanding out superpotential (\ref{eq:W_np}) as a power series in ${\cal Z}_i$, one can show that the coefficient of unnormalized quartic term comes out to be \cite{ferm_masses_MS}:
{\small
$$\hskip-0.1in{\cal O}_{{\cal Z}_i{\cal Z}_j{\cal Z}_k{\cal Z}_l}\sim \frac{2^{n^s}}{24}10^2(\mu_3 n^s l^2)^4{\cal V}^{\frac{n^s}{2}+\frac{1}{9}}
e^{-n^s {\rm vol}(\Sigma_S) + i n^s \mu_3l^2{\cal V}^{\frac{1}{18}}(\alpha+i\beta)},$$
}
 where $\alpha,\beta\sim{\cal O}(1)$ constants and $l=2\pi\alpha^\prime$.

Now, we will elaborate on running of the neutrino mass.
Now, as shown in \cite{Babu_et_al}, unlike MSSM, there are multiple dimension-five operators
in 2HDM corresponding to the Higgses. In our setup, we have taken the two Higgses to be on the same footing
and hence along this locus, there is only one dimension-five operator in the 2HDM as well. In
this (and the LVS) limit(s), and assuming that the Higgses couple only to the $u$-type quarks
as well as taking the $U(1)_Y$ fine structure constant to be equal to the coefficient of the term quartic in the Higgs\footnote{Given that $\lambda\sim(n^s\mu_3l^2)^4\sim1/\pi^4$ in our setup, this would imply, e.g., at the string scale $g_{U(1)}^2\sim0.01$, which is quite reasonable. See also
K.~Sasaki, M.~S.~Carena and C.~E.~M.~Wagner,
  Nucl.\ Phys.\  B {\bf 381}, 66 (1992), for justification.}, one then sees that the 2HDM and MSSM RG flow equations for $\kappa$ become identical. Using then the one-loop solution to the $\langle H_u\rangle$ RG flow equation for the 2HDM as given in \cite{Das_Parida}, one obtains:
\begin{eqnarray}
\label{eq:sol_2HDM_Hu_I}
& & (v sin\beta)_{M_s}=\langle H_u\rangle_{M_s} =\langle H_u\rangle_{M_{EW}}e^{-\frac{3}{16\pi^2}\int_{ln(M_{EW})}^{ln(M_S)}Y^2_t dt^\prime}\nonumber\\
& & \hskip-0.3in\times(\alpha_1(M_s)/\alpha(M_{EW}))^{3/56}(\alpha_2(M_s)/\alpha_2(M_{EW}))^{-3/8},
\end{eqnarray}
where using arguments of the next paragraph, one can set the exponential to unity.
Now, in the dilute flux approximation, the coupling-constants-dependent factor is
$(\alpha_2(M_s)/\alpha_2(M_{EW}))^{-9/28}$. Further, using the one-loop RG flow solution for
$\alpha_2$ of \cite{Sparticles_Misra_Shukla}:
\begin{equation}
\label{eq:sol_2HDM_Hu_II}
(v sin\beta)_{M_{EW}}\sim(v sin\beta)_{M_s}\left(1-{\cal O}(60)/\frac{16\pi^2}{g^2(M_s)}\right)^{\frac{9}{28}},
\end{equation}
where $(v sin\beta)_{M_{EW}}\sim246 GeV$ in the large $tan\beta$ regime. Hence, by requiring
$g^2(M_s)$ to be sufficiently close to $16\pi^2/{\cal O}(60)\sim2.5$, one can RG flow $\langle H_u\rangle_{M_s}$ to the required value $\langle H_u\rangle_{M_{EW}}\sim 246 GeV$.

The analytic solution to RG running equation for $\kappa$ is given by:
$$\kappa(M_s)
=\frac{\kappa_{\tau\tau}(M_s)}{\kappa_{\tau\tau}(M_{EW})}\left(\begin{array}{ccc}
\frac{I_e}{I_\tau}& 0 & 0\\
0 & \frac{I_\mu}{I_\tau} & 0 \\
0 & 0 & 1
\end{array}\right)\kappa(M_{EW})\left(\begin{array}{ccc}
\frac{I_e}{I_\tau}& 0 & 0\\
0 & \frac{I_\mu}{I_\tau} & 0 \\
0 & 0 & 1
\end{array}\right)$$
where $I_{e/\mu/\tau}=e^{\frac{1}{8\pi^2}\int_{ln M_{EW}}^{ln M_S}dt\hat{Y}_{e/\mu/\tau}^2}$. Now, for $tan\beta=\frac{\langle z_1\rangle}{\langle z_2\rangle}<50$,
$\frac{I_{e/\mu}}{I_\tau}\approx \left(1-\frac{\hat{Y}_\tau^2}{8\pi^2}ln\left(\frac{M_S}{M_{EW}}\right)\right)$
\cite{RG_neutrino_I}. Assuming $\hat{Y}_t^2/(4\pi)^2\sim10^{-5}$ - see \cite{Ibanez_et_al} - one sees that $I_{e/\mu/\tau}\approx 1$. As $\kappa_{\tau\tau}$ is an overall factor in $\tau$, one can argue that it can be taken to be scale-independent \cite{energy_MNS}. Hence, in MSSM and 2HDM, the coefficient of quartic term $\hat{{\cal O}}_{{\cal Z}_i{\cal Z}_i{\cal Z}_i{\cal Z}_i}$ does not run. For ${\cal V}\sim10^6$ (in string length units), $n^s=2$  and reduced Planck mass $M_p=2.4\times10^{18}{\rm GeV}$,  one obtains:
\begin{equation}
 \label{eq:final_I}
 m_\nu=\frac{(v sin\beta)_{M_{EW}}^2\hat{O}_{{\cal Z}_i{\cal Z}_j{\cal Z}_k{\cal Z}_l}}{2M_p}\stackrel{<}{\sim} 1eV.
 \end{equation}

 \section{Conclusion}

 We reviewed our setup involving a mobile space-time filling $D3$-brane and stack(s) of fluxed $D7$-brane(s) wrapping the ``big" divisor in the context of type IIB Swiss-Cheese Calabi-Yau orientifolds,  to obtain a variety of results such as ${\cal O}(1)\ g_{YM}$, the soft supersymmetry breaking parameters including the soft scalar masses (and their RG-flow up to one loop), the physical Yukawa couplings, the physical Higgsino mass parameter as well as fermionic mass scales corresponding to the first two generations of quarks/leptons and neutrinos (and their RG-flow up to one loop). Very heavy scalars and light fermions indicate the possibility of a split-SUSY large volume Swiss-Cheese scenario\cite{Mansi_Aalok}. With the inclusion of only a mobile space-time filling $D3$-brane, we also reviewed our proposal of obtaining a very heavy gravitino in the inflationary epoch and a light gravitino in the present era, within the same string theoretic setup.

 \section*{Acknowledgement}

 It is a pleasure to thank Pramod Shukla for collaboration on all issues discussed in sections {\bf 2} and {\bf 3}. This work was partly supported by a junior associateship at the Abdus Salam ICTP. In addition, the author would like to thank for their hospitality and discussions, the string theory groups at Harvard, Institute of Advanced Study (Princeton), McGill, Imperial, Enrico Fermi Institute (U.Chicago), Maryland Center for Fundamental Physics, Max Planck Institute for Physics at Munich and the Albert Einstein Institute for Gravitation at Golm where, since 2009, different stages of the work summarized in this review, were completed.

\end{document}